\documentclass{article}
\usepackage{amsmath}
\usepackage{mathtools}
\usepackage{amssymb}
\usepackage{graphicx}
\usepackage{caption}
\usepackage{subcaption}
\title{Two entanglement conditions and their connection to negativity}
\author{Mark Hillery$^{1,2}$, Camilla Polvara$^{2,3}$, \\ Vadim Oganesyan$^{2,3,4,5}$, and Nada Ali$^{1,2}$ \\ \\
$^{1}$ Department of Physics and Astronomy, Hunter College of CUNY, \\
695 Park Avenue, New York, NY 10065 \\
$^{2}$ Physics Program, Graduate Center of CUNY, \\ 365 Fifth Avenue, New York, NY 10016 \\
$^{3}$ Department of Physics and Astronomy, \\ College of Staten Island (CUNY),  \\ 2800 Victory Blvd. , Staten Island, NY 10314 \\ 
$^{4}$ Initiative in the Theoretical Sciences, Graduate Center of CUNY, \\ 365 Fifth Avenue, New York, NY 10016 \\
$^{5}$ Center for Computational Quantum Physics, Flatiron Institute \\ 162 Fifth Avenue , New York, NY 10010}
\begin{document}
\maketitle

\begin{abstract}
We examine two conditions that can be used to detect bipartite entanglement, and show that they can be used to provide lower bounds on the negativity of states.  We begin with two-qubit states, and then show how what was done there can be extended to more general states.  The resulting bounds are then studied by means of a number of examples.  We also show that if one has some knowledge of the Schmidt vectors of a state, better bounds can be found.
\end{abstract}

\section{Introduction}
Entanglement is a resource for many different tasks in quantum information, among them, teleportation, entanglement swapping, and some forms of quantum cryptography.  Quantifying the entanglement in a state can then be a useful in determining how much of this resource a given procedure will use.  For bipartite entanglement, the von Neumann entropy of the reduced density matrix of the state is the standard measure of entanglement.  It can, however, sometimes be difficult to calculate, in particular for mixed states of large dimension.  An alternative, which can be useful for bipartite states whose partial transpose is not positive is the negativity \cite{z1,z2,werner}.  It was first defined in \cite{z1}, used to study the volume of separable states in \cite{z2}, and a number of its properties analyzed in \cite{werner}.  For large systems finding the negativity can also be laborious, and in some cases, all one is interested in is a guarantee that at least a certain amount of entanglement is present.  This can often be accomplished with simple conditions \cite{cavalcanti,plenio,eisert,guhne}, for example by making a small number of measurements or using entanglement witnesses.

Here we want to see if we can use some other conditions that show the existence of entanglement to place bounds on how entangled a state is.  It was shown that if we have a state in $\mathcal{H}=\mathcal{H}_{a} \otimes \mathcal{H}_{b}$, that the state is entangled if
\begin{eqnarray}
\label{inequalities}
|\langle A^{\dagger}B\rangle |^{2} & > & \langle A^{\dagger}AB^{\dagger}B\rangle \nonumber  \\
& {\rm or} & \nonumber \\
|\langle AB\rangle |^{2} & > & \langle A^{\dagger}A\rangle \langle B^{\dagger}B\rangle ,
\end{eqnarray}
where $A$ and $B$ are arbitrary non-hermitian operators, and $A$ acts in $\mathcal{H}_{a}$ and $B$ acts in $\mathcal{H}_{b}$ \cite{hillery1,hillery2}.  If for a particular state, neither of these conditions is satisfied, then we can say nothing about the entanglement of that state.  We should also note that these conditions will detect entanglement in a state only if the partial transpose of the state is not positive \cite{hillery3}.  We would like to see if we can relate the extent to which one of these inequalities is satisfied, i.e.\ the difference between the left and right-hand sides of the inequality, to a measure of the entanglement of a state.  Since these conditions only detect entanglement in states whose partial transpose is not positive, the natural measure to study is the negativity.  

The negativity of a state is defined as follows.  Let $\rho$ be a state on $\mathcal{H}_{a}\otimes \mathcal{H}_{b}$, and let $\rho^{T_{b}}$ be the partial transpose of $\rho$ with respect to system $b$, i.e.
\begin{equation}
\rho^{T_{b}}_{m,\mu ; n,\nu} = \,_{a}\langle m|\,_{b}\langle \mu |\rho^{T_{b}}|n\rangle_{a} |\nu \rangle_{b} = \,_{a}\langle m|\,_{b}\langle \nu |\rho |n\rangle_{a} |\mu \rangle_{b} = \rho_{m,\nu ; n , \mu} .
\end{equation}
The negativity of $\rho$ is defined to be
\begin{equation}
N(\rho )= \frac{\| \rho^{T_{b}}\|_{1} -1}{2} ,
\end{equation}
where the above norm is the trace norm.  The idea behind this measure is that if the partial transpose of a state is not positive, it is entangled.  If the partial transpose of a state is positive, then $\| \rho^{T_{b}}\|_{1} ={\rm Tr}(\rho^{T_{b}}) =1$, and the negativity is zero.  On the other hand, suppose
\begin{equation}
\rho^{T_{b}} = \sum_{j} \lambda_{j}^{(+)} |\phi^{(+)}_{j}\rangle\langle\phi^{(+)}| + \sum_{j} \lambda_{j}^{(-)}
|\phi^{(-)}_{j}\rangle\langle\phi^{(-)}| ,
\end{equation}
where $\lambda_{j}^{(+)}>0$ and $\lambda_{j}^{(-)}<0$.  Now, because the trace of the transpose of an operator
is the same as the trace of the operator, we have that
\begin{equation}
\sum_{j} \lambda_{j}^{(+)} + \sum_{j} \lambda_{j}^{(-)} =1 ,
\end{equation}
but
\begin{equation}
\| \rho^{T_{b}}\|_{1} = \sum_{j} \lambda_{j}^{(+)} + \sum_{j} |\lambda_{j}^{(-)}| >1 .
\end{equation}
Thus, negativity measures how many and how big the negative eigenvalues in the partial transpose are.
Negativity is convex and monotonic under probabilistic LOCC \cite{werner}.

The quantities in Eq.\ (\ref{inequalities}) can be measured, since any non-hermitian operator, $A$, can be expressed in terms of its real and imaginary parts, $(A+A^{\dagger})/2$ and $i(A-A^{\dagger})/2$, which are hermitian.  If these operators are simple, for example rank one projections, they should not be too hard to measure.  The information gained from such measurements, while no substitute for full state tomography, can provide a floor to the negativity of a state.  Determining the actual negativity does require state tomography, and requires far more measurements that the procedures outlined here would entail.
 
\section{Two qubits}
We will first look at a very simple example, and then see what can be done in a more general setting.  Let
us start with the first condition.  We will consider two qubits, which we shall denote by $a$ and $b$, and set $A=\sigma_{a}^{(-)}$ and $B=\sigma_{b}^{(-)}$, where $\sigma_{a}^{(-)}=|0\rangle_{a}\langle 1|$ and $\sigma^{(-)}_{b}=|0\rangle_{b}\langle 1|$.  We then have that
\begin{equation}
\langle A^{\dagger}B\rangle = \langle \sigma_{a}^{(+)}\sigma_{b}^{(-)}\rangle = \rho_{01;10} ,
\end{equation}
where we made use of the fact that $\sigma^{(+)}_{a}=|1\rangle_{a}\langle 0|$.  We
also have that
\begin{equation}
\langle  A^{\dagger}AB^{\dagger}B\rangle = \langle \sigma^{(+)}_{a}\sigma_{a}^{(-)}\sigma_{b}^{(+)}
\sigma_{b}^{(-)}\rangle = \rho_{11;11} .
\end{equation}
Expressing all quantities in terms of matrix elements of the partially transposed density matrix, we have that
$\rho^{T_{B}}_{00;11}=\rho_{01;10}$ and $\rho^{T_{B}}_{11;11}=\rho_{11;11}$, so our first entanglement condition
becomes
\begin{equation}
\label{Mcondition}
\kappa = |\rho^{T_{B}}_{00;11}|^{2} - \rho^{T_{B}}_{11;11} > 0.
\end{equation}
Now the matrix
\begin{equation}
M=\left( \begin{array}{cc} \rho^{T_{B}}_{00;00} & \rho^{T_{B}}_{00;11} \\ \rho^{T_{B}}_{11;00} & \rho^{T_{B}}_{11;11}
\end{array} \right)  ,
\end{equation}
is a sub-matrix of the $4\times 4$ matrix $\rho^{T_{B}}$.  Now, let $\tilde{v}$ be a normalized four-component 
vector in the two-qubit space whose $\tilde{v}_{01}$ and $\tilde{v}_{10}$ components are zero, that is only 
$\tilde{v}_{00}$ and $\tilde{v}_{11}$ are nonzero.  Then
\begin{equation}
\langle \tilde{v}| \rho^{T_{B}}| \tilde{v}\rangle = (\tilde{v}_{00}^{\ast}\ \tilde{v}_{11}^{\ast} ) M\left( \begin{array}{c} 
\tilde{v}_{00} \\ \tilde{v}_{11} \end{array} \right)  .
\end{equation}
For a general normalized four-component vector $v$, the smallest $\langle v| \rho^{T_{B}}|v\rangle$ can be is
the lowest negative eigenvalue of $\rho^{T_{B}}$, which we shall denote by $\lambda_{min}$.  The above
equation implies that
\begin{equation}
\lambda_{min} \leq (\tilde{v}_{00}^{\ast}\ \tilde{v}_{11}^{\ast} ) M\left( \begin{array}{c} \tilde{v}_{00} \\ \tilde{v}_{11} \end{array} \right) ,
\end{equation}
and this further implies that the lowest eigenvalue of $M$, which is the smallest value the right-hand side can
attain, must be greater than $\lambda_{min}$.

Next we want to see what the condition in Eq.\ (\ref{Mcondition}) places on the eigenvalues of $M$.  We have that
\begin{equation}
\det M =  \rho^{T_{B}}_{00;00} \rho^{T_{B}}_{11;11} - | \rho^{T_{B}}_{00;11}|^{2} \leq -\kappa ,
\end{equation}
where we have used the fact that $0\leq \rho^{T_{B}}_{00;00},\ \rho^{T_{B}}_{11;11} \leq 1$.  The fact that
$\det M<0$ means that one of the eigenvalues is positive and the other is negative.  Let the positive eigenvalue
be $\xi_{+}$ and the negative one be $\xi_{-}$.  We then have that
\begin{equation}
\xi_{+} + \xi_{-} \leq 1 \hspace{5mm} \xi_{+}\xi_{-} \leq -\kappa .
\end{equation}
If we now plot the regions allowed by these two conditions, keeping in mind that $\xi_{+}>0$ and $\xi_{-}<0$,
we see that the largest possible value of $\xi_{-}$ is given by the point where the line $\xi_{-}=1-\xi_{+}$ and
the branch of $ \xi_{+}\xi_{-} \leq -\kappa $ with $\xi_{+}>0$ and $\xi_{-}<0$ intersect.  This happens when
\begin{equation}
1-\xi_{+}= \frac{-\kappa}{\xi_{+}} .
\end{equation}
This gives
\begin{equation}
\xi_{+}=\frac{1}{2}[ 1+ (1+4\kappa )^{1/2}] \hspace{5mm} \xi_{-}=1-\xi_{+} = \frac{1}{2}[ 1- (1+4\kappa )^{1/2}] .
\end{equation}
We, therefore, have that 
\begin{equation}
\lambda_{min} \leq \frac{1}{2}[ 1- (1+4\kappa )^{1/2}]  .
\end{equation}
Combining this result with 
\begin{equation}
N(\rho ) = \frac{1}{2} \sum_{j=1}^{4} (|\lambda_{j}|-\lambda_{j}) \geq \frac{1}{2} (|\lambda_{min}|-\lambda_{min}) = -\lambda_{min} ,
\end{equation}
we find that
\begin{equation}
\label{qubit}
N(\rho ) \geq  \frac{1}{2}[(1+4\kappa )^{1/2} - 1] .
\end{equation}
This is the desired result that gives us a lower bound on the negativity in terms of $\kappa$, a quantity that is
directly related to out entanglement condition.

It is possible to improve this result if we have more information.  In particular, if we know $\rho^{T_{B}}_{00;00} + \rho^{T_{B}}_{11;11} = \rho_{00;00} + \rho_{11;11}$, which we shall denote by $a$, we can replace the condition $\xi_{+} + \xi_{-} \leq 1$ by $\xi_{+} + \xi_{-} \leq a$.  This leads to the condition
\begin{equation}
\xi_{-} = \frac{1}{2} [ a - (a^{2} + 4\kappa )^{1/2} ] .
\end{equation}
and 
\begin{equation}
N(\rho ) \geq \frac{1}{2} [ (a^{2}+4\kappa )^{1/2} - a ] .
\end{equation}
Let's call the right-hand side $g(a)$ and look at it as a function of $a$ for $0 \leq a \leq 1$.  We see that
\begin{equation}
\frac{dg}{da} = \frac{1}{2} \left[ \frac{a}{(a^{2}+4\kappa )^{1/2}} - 1\right] < 0 ,
\end{equation}
which means that $g(a)$ is a decreasing function of $a$.  Therefore, setting $a= \rho_{00;00} + \rho_{11;11}$ instead of $a=1$ will produce a better inequality.

\section{Extending the two-qubit result}

We can extend the result in Eq.\ (\ref{qubit}) to systems beyond qubits.  Suppose we have a bipartite state on $\mathcal{H}_{a}\otimes\mathcal{H}_{b}$, where $\mathcal{H}_{a}$ and $\mathcal{H}_{b}$ can be of any finite dimension.  Let $A=|\eta_{0}\rangle_{a}\langle \eta_{1}|$ and $B=|\xi_{0}\rangle_{b}\langle\xi_{1}|$, where $\langle\eta_{0}|\eta_{1}\rangle = 0$, and $\langle\xi_{0}|\xi_{1}\rangle = 0$.  If we make the identifications $|j\rangle_{a} \leftrightarrow |\eta_{j}\rangle_{a}$ and $|j\rangle_{b} \leftrightarrow |\xi_{j}\rangle_{b}$ for $j=0,1$, then the arguments in the qubit section go through.  Another way of doing this is to note that for any projection operator $\Pi_{ab}$ on $\mathcal{H}_{a}\otimes\mathcal{H}_{b}$, we have $\|\rho^{T_{B}}\|_{1} \geq \|\Pi_{ab}\rho^{T_{B}}\Pi_{ab} \|_{1}$ and we can choose $\Pi_{ab}=P_{1}^{(a)} \otimes P_{1}^{(b)}$, where $P_{1}^{(a)} = |\eta_{0}\rangle_{a}\langle\eta_{0}| + |\eta_{1}\rangle_{a}\langle\eta_{1}|$ and $P_{1}^{(b)} = |\xi_{0}\rangle_{b}\langle\xi_{0}| + |\xi_{1}\rangle_{b}\langle\xi_{1}|$.  The matrix for $\Pi_{ab}\rho^{T_{B}}\Pi_{ab}$ is $4\times 4$ and equivalent to two qubits.

For the higher dimensional systems, we can improve our results.  Suppose we have two sets of operators, $A_{1}$ and $B_{1}$, which are as before, and $A_{2}=|\eta_{2}\rangle_{a}\langle \eta_{3}|$ and $B=|\xi_{2}\rangle_{b}\langle\xi_{3}|$, where $\{  |\eta_{j}\rangle \, |\, j=0,1,2,3\}$ and $\{  |\xi_{j}\rangle \, |\, j=0,1,2,3\}$ are orthonormal sets.  Let us further suppose that for the state $\rho$, 
\begin{equation}
|\langle A_{j}^{\dagger}B_{j}\rangle |^{2} - \langle A_{j}^{\dagger}A_{j}B_{j}^{\dagger}B_{j}\rangle =  \kappa_{j} > 0 ,
\end{equation}
for $j=1,2$.  We can now make use of the following result, which will be proved shortly.  If $\Pi_{ab}^{(1)}$ and $\Pi_{ab}^{(2)}$ are two orthogonal projections, i.e.\ $\Pi_{ab}^{(1)} \Pi_{ab}^{(2)} = 0$, then
\begin{equation}
\label{proj}
\|\rho^{T_{B}}\|_{1} \geq \|\Pi_{ab}^{(1)}\rho^{T_{B}}\Pi_{ab}^{(1)} \|_{1} + \|\Pi_{ab}^{(2)}\rho^{T_{B}}\Pi_{ab}^{(2)} \|_{1} .
\end{equation}
In our case we choose $\Pi_{ab}^{(1)}= P_{1}^{(a)} \otimes P_{1}^{(b)}$ as before and $\Pi_{ab}^{(2)} = P_{2}^{(a)} \otimes P_{2}^{(b)}$, where   $P_{2}^{(a)} = |\eta_{2}\rangle_{a}\langle\eta_{2}| + |\eta_{3}\rangle_{a}\langle\eta_{3}|$, and $P_{2}^{(b)} = |\xi_{2}\rangle_{b}\langle\xi_{2}| + |\xi_{3}\rangle_{b}\langle\xi_{3}|$.  We can then apply Eq.\ (\ref{qubit}) twice to get
\begin{equation}
\label{extend}
N(\rho ) \geq  \frac{1}{2}\sum_{j=1}^{2} [(1+4\kappa_{j} )^{1/2} - 1] .
\end{equation}

We now need to prove Eq.\ (\ref{proj}).  We start by noting that any hermitian operator can be expressed as the difference of two positive operators, in particular $\rho^{T_{B}} = S - T$, where $S$ and $T$ are positive.  This implies that $\| \rho^{T_{B}}\|_{1} = {\rm Tr}(S+T)$.  Now define
\begin{eqnarray}
\rho_{p}^{T_{B}} & = & \Pi_{ab}^{(1)}\rho^{T_{B}}\Pi_{ab}^{(1)} + \Pi_{ab}^{(2)}\rho^{T_{B}}\Pi_{ab}^{(2)} \nonumber \\
& = & \Pi_{ab}^{(1)}S\Pi_{ab}^{(1)} - \Pi_{ab}^{(1)}T\Pi_{ab}^{(1)} + \Pi_{ab}^{(2)}S\Pi_{ab}^{(2)} - \Pi_{ab}^{(2)}T\Pi_{ab}^{(2)} .
\end{eqnarray}
Now this equation expresses $\rho_{p}^{T_{B}}$ as the difference of two positive operators, so
\begin{eqnarray}
\| \rho_{p}^{T_{B}}\|_{1} & = & {\rm Tr}( \Pi_{ab}^{(1)}S\Pi_{ab}^{(1)} + \Pi_{ab}^{(1)}T\Pi_{ab}^{(1)} + \Pi_{ab}^{(2)}S\Pi_{ab}^{(2)} + \Pi_{ab}^{(2)}T\Pi_{ab}^{(2)}) \nonumber \\
& = & \|  \Pi_{ab}^{(1)}\rho^{T_{B}}\Pi_{ab}^{(1)}\|_{1} + \|  \Pi_{ab}^{(2)}\rho^{T_{B}}\Pi_{ab}^{(2)} \|_{1} .
\end{eqnarray}
But we also have that 
\begin{eqnarray}
{\rm Tr}(S+T) & = & \|  \rho^{T_{B}}\|_{1} \geq {\rm Tr}( \Pi_{ab}^{(1)}(S+T)\Pi_{ab}^{(1)} +  \Pi_{ab}^{(2)}(S+T)\Pi_{ab}^{(2)} ) ,
\end{eqnarray}
so that $\| \rho^{T_{B}}\|_{1} \geq \| \rho_{p}^{T_{B}}\|_{1} $.

\section{A second approach}

Using different methods it is possible to derive a different negativity bound from the first inequality.  We define $\kappa$ as before and wish to find a lower bound on the negativity in terms of it.  We shall assume that the operators $A$ and $B$ are bounded.  Note
that this means that the case in which they they are mode annihilation operators is not covered here, because
annihilation operators are unbounded.  We begin by noting that
\begin{eqnarray}
{\rm Tr}(\rho AB^{\dagger})={\rm Tr}(\rho^{T_{B}}A(B^{\dagger})^{T_{B}})= {\rm Tr}(\rho^{T_{B}}AB^{\ast}) 
\nonumber \\
{\rm Tr}(\rho A^{\dagger}AB^{\dagger}B)={\rm Tr}(\rho^{T_{B}} A^{\dagger}A(B^{\dagger}B)^{\ast}) .
\end{eqnarray}
Now $\rho^{T_{B}}$ can be expressed as the difference of two positive operators, $\rho_{+}$ and $\rho_{-}$,
\begin{equation}
\rho^{T_{B}}=\rho_{+} -\rho_{-} .
\end{equation}
Letting $\mu_{\pm}={\rm Tr}(\rho_{\pm})$, we have that $\mu_{+}-\mu_{-}=1$ and $N(\rho )=(\mu_{+}+\mu_{-} -1)/2 = \mu_{-}$.

Denote the eigenvalues and eigenstates of $\rho_{+}$ as $\lambda^{(+)}$ and $|\psi_{n}^{(+)}\rangle$, 
respectively.  We then have that
\begin{eqnarray}
|{\rm Tr}(\rho_{+}AB^{\ast})| &=& |\sum_{n} \lambda_{n}^{(+)}\langle\psi_{n}^{(+)}|AB^{\ast}|\psi_{n}^{(+)}\rangle |
\leq \sum_{n} \lambda_{n}^{(+)}(\langle \psi_{n}^{(+)}|A^{\dagger}A(B^{\ast})^{\dagger}B^{\ast}|\psi_{n}^{(+)}
\rangle )^{1/2}  \nonumber \\
& \leq & (\sum_{n} \lambda_{n}^{(+)})^{1/2} (\sum_{n} \lambda_{n}^{(+)} \langle \psi_{n}^{(+)}|A^{\dagger}A
(B^{\ast})^{\dagger}B^{\ast}|\psi_{n}^{(+)} \rangle )^{1/2} \nonumber \\
& \leq & \sqrt{\mu_{+}} [ {\rm Tr}(\rho_{+} A^{\dagger}A (B^{\dagger}B))^{\ast} ]^{1/2} ,
\end{eqnarray}
were we have used the fact that $(B^{\ast})^{\dagger}B^{\ast}=(B^{\dagger}B)^{\ast}$.

Now define $z_{\pm}={\rm Tr}(\rho_{\pm}AB^{\ast})$ and $V_{\pm}={\rm Tr}(\rho_{\pm}A^{\dagger}A(B^{\dagger}
B)^{\ast})$.  Note that
\begin{equation}
z_{+}-z_{-}=\langle AB^{\dagger}\rangle \hspace{5mm} V_{+}-V_{-}=\langle A^{\dagger}AB^{\dagger}B\rangle .
\end{equation}
In addition, from above we have that $|z_{+}|\leq (\mu V_{+})^{1/2}$, and, similarly, $|z_{-}|\leq (\mu V_{-})^{1/2}$.
Therefore, we have that,
\begin{equation}
\kappa = |z_{+}-z_{-}|^{2} - \langle A^{\dagger}AB^{\dagger}B\rangle \leq (|z_{+}| + |z_{-}|)^{2} 
- \langle A^{\dagger}AB^{\dagger}B\rangle ,
\end{equation}
and this implies that
\begin{eqnarray}
\kappa + \langle A^{\dagger}AB^{\dagger}B\rangle & \leq & [ (\mu_{+}V_{+})^{1/2} + (\mu_{-}V_{-})^{1/2} ]^{2}
\nonumber \\
& \leq & [ (1+\mu_{-})^{1/2}(V_{-} + \langle A^{\dagger}AB^{\dagger}B\rangle )^{1/2} + (\mu_{-}V_{-})^{1/2} ]^{2} .
\end{eqnarray}
We now make use of the assumption that the operators $A$ and $B$ are bounded.  We have that
\begin{equation}
|V_{-}|\leq \| \rho_{-}A^{\dagger}A(B^{\dagger}B)^{\ast}\|_{1} \leq \mu_{-} \| A^{\dagger}A(B^{\dagger}B)^{\ast}\| ,
\end{equation}
where the norm with the subscript $1$ is the trace norm and the norm without a subscript is the operator norm.
Substituting this into the above inequality we find that
\begin{eqnarray}
\label{method2}
[\kappa + \langle A^{\dagger}AB^{\dagger}B\rangle ]^{1/2} & \leq &  (1+\mu_{-})^{1/2} [\mu_{-} \| A^{\dagger}A
(B^{\dagger}B)^{\ast}\| + \langle A^{\dagger}AB^{\dagger}B\rangle ]^{1/2} \nonumber \\
& & +\mu_{-}\| A^{\dagger}A(B^{\dagger}B)^{\ast}\|^{1/2} .
\end{eqnarray}
This inequality should give us a lower bound for $\mu_{-}$ in terms of $\kappa$, and because $N(\rho)=\mu_{-}$ it will give us a lower bound on the negativity.  Note that for $\kappa = 0$, we have that 
$\mu_{-} =0$ satisfies the inequality, so that presumably, if $\kappa > 0$, we will get a non-zero lower bound
for $\mu_{-}$.  In addition, note that $| A^{\dagger}A(B^{\dagger}B)^{\ast}\| = \| A^{\dagger}A\|\, \|(B^{\dagger}B)^{\ast}\|$.  Now $(B^{\dagger}B)^{\ast}$ is obtained by taking the matrix for $B^{\dagger}B$ in the computational basis (operator transpose and complex conjugation are basis dependent operations) and taking its complex conjugate.  This implies that the eigenvalues of $(B^{\dagger}B)^{\ast}$ are just the complex conjugates of the eigenvalues of $B^{\dagger}B$, so that $\| (B^{\dagger}B)^{\ast}\| = \| B^{\dagger}B \|$.

We can obtain a slightly simpler version of this inequality by making use of the inequality
\begin{equation}
(1+ z)^{1/2} = 1 + \int_{0}^{z} ds\, \frac{1}{2} \frac{1}{(1+s)^{1/2}} \leq 1+\frac{z}{2} ,
\end{equation}
to simplify the right-hand side of the inequality.  Setting  $x=\langle A^{\dagger}AB^{\dagger}B\rangle$ and $y=\| A^{\dagger}A(B^{\dagger}B)^{\ast}\|$, we have that
\begin{equation}
(\kappa + x)^{1/2} \leq \left( 1 + \frac{\mu_{-}}{2} \right) \sqrt{x} \left( 1 + \frac{y\mu_{-}}{2x} \right) + \mu_{-}\sqrt{y} .
\end{equation}
This can be easily solved for $\mu_{-}$.

\section{Second inequality for qubits}
Let's look at the second inequality in Eq.\ (\ref{inequalities}) for two qubits, and choose $A=\sigma_{a}^{(+)}$ and $B=\sigma_{b}^{(+)}$.  In terms of density matrix elements, we have
\begin{eqnarray}
\langle AB\rangle & = & \rho_{00;11} = \rho^{T_{B}}_{01;10} \nonumber \\
\langle A^{\dagger}A\rangle & = & \sum_{j=0}^{1} \rho_{0j;0j} = \sum_{j=0}^{1} \rho^{T_{B}}_{0j;0j} \nonumber \\
\langle B^{\dagger}B\rangle & = & \sum_{j=0}^{1}\rho_{j0;j0} = \sum_{j=0}^{1}\rho^{T_{B}}_{j0;j0} .
\end{eqnarray}
Substituting these expressions into the second inequality we find for the difference of the two sides
\begin{equation}
\label{condition2}
\kappa = |\langle AB\rangle |^{2} - \langle A^{\dagger}A\rangle\langle B^{\dagger}B\rangle = |\rho^{T_{B}}_{01;10} |^{2} - ( \rho^{T_{B}}_{00;00} + \rho^{T_{B}}_{01;01} )(\rho^{T_{B}}_{00;00} + \rho^{T_{B}}_{10;10} ) .
\end{equation}
Now $\rho^{T_{B}}_{01;01}$, $\rho^{T_{B}}_{10;10}$,  $\rho^{T_{B}}_{01;10}$, and its complex conjugate form a $2\times 2$ matrix, which we shall denote by $M$.  By the same argument as before, the absolute value of the lowest negative eigenvalue of this $2\times 2$ matrix will be a lower bound for the negativity.  The difference from the previous case is that the above expression involves one additional quantity besides the elements of the $2\times 2$ matrix, $\rho^{T_{B}}_{00;00}$, which we shall denote, for convenience, by $1\geq \alpha \geq 0$.

Let the eigenvalues of $M$ be $\xi_{1}$ and $\xi_{2}$.  Since ${\rm det}(M)=\xi_{1}\xi_{2}$ and $\rho^{T_{B}}_{01;01} + \rho^{T_{B}}_{10;10} = \xi_{1}+\xi_{2}$, we find
\begin{equation}
\label{hyp}
-\kappa = (\xi_{1} + \alpha )(\xi_{2} + \alpha ) .
\end{equation}
We also have the condition, since ${\rm Tr}(\rho^{T_{B}}) = 1$, that $\alpha + \xi_{1}+\xi_{2} \leq 1$.  In order to satisfy the first condition, we must have either $\xi_{1} + \alpha < 0$ or $\xi_{2} + \alpha < 0$.  Let's assume $\xi_{2} + \alpha < 0$.  The next step is to find where the line $\alpha + \xi_{1}+\xi_{2} = 1$ intersects the hyperbola in Eq.\ (\ref{hyp}).  Solving both equations for $\xi_{1}$ and then putting them equal to each other we find
\begin{equation}
1 - \alpha - \xi_{2} = \frac{-\kappa}{\xi_{2} + \alpha} - \alpha ,
\end{equation}
with the negative solution
\begin{equation}
\xi_{2} = \frac{1}{2} \{ 1-\alpha - [(1+\alpha )^{2} + 4\kappa]^{1/2} \} ,
\end{equation}
which implies that
\begin{equation}
\xi_{2} \leq \frac{1}{2} \{ 1-\alpha - [(1+\alpha )^{2} + 4\kappa]^{1/2} \} .
\end{equation}
This implies that the negativity must satisfy
\begin{equation}
N(\rho ) \geq \frac{1}{2} \{ [(1+\alpha )^{2} + 4\kappa]^{1/2} -1 + \alpha \} .
\end{equation}
Note that even when $\kappa = 0$, the bound for the negativity is not zero.  That means that the condition that $\xi_{2} < 0$ is a stronger condition for detecting entanglement than the condition $\kappa > 0$, where $\kappa$ is given by Eq.\ (\ref{condition2}).

\section{Examples}
For a bipartite state expressed in terms of its Schmidt basis
\begin{equation}
|\Psi\rangle = \sum_{j=1}^{N} \sqrt{\lambda_{k}}|u_{j}\rangle_{a} |v_{j}\rangle_{b} ,
\end{equation}
the negativity is \cite{werner}
\begin{equation}
N(|\Psi\rangle_{ab}\langle\Psi |) = \frac{1}{2} \left[ \left( \sum_{j=1}^{N} \sqrt{\lambda_{j}} \right)^{2} -1 \right] .
\end{equation} 
We can use this to study several simple examples and compare the actual negativity of a state to the lower bounds we have obtained.

Consider the two-qubit state $|\Psi\rangle_{ab} = \sqrt{\lambda_{0}} |01\rangle_{ab} + \sqrt{\lambda_{1}} |10\rangle_{ab}$ and let $A=\sigma^{(-)}_{a}$, and $B=\sigma^{(-)}_{b}$.  We then have that $\langle A^{\dagger}AB^{\dagger}B\rangle = 0$ and $\langle A^{\dagger}B\rangle = \sqrt{\lambda_{0}\lambda_{1}}$ giving us $\kappa =  \lambda_{0}\lambda_{1}$.  Substituting this into Eq.\ (19) gives us
\begin{equation}
N(\rho )\geq \frac{1}{2} [ (1+4\lambda_{0}\lambda_{1})^{1/2}-1] .
\end{equation}
Now in this case we know that $N(\rho )= \sqrt{\lambda_{0}\lambda_{1}}$, so let's first verify the inequality.  Substituting and rearranging we get
\begin{equation}
\sqrt{\lambda_{0}\lambda_{1}} + \frac{1}{2}  \geq \left( \frac{1}{4} + \lambda_{0}\lambda_{1}\right)^{1/2} ,
\end{equation}
which can be seen to be true by squaring both sides. Now $0\leq \lambda_{0}\lambda_{1} \leq 1/4$, and when $\lambda_{0}\lambda_{1} = 0$, both the negativity and the bound are $0$, whereas when $\lambda_{0}\lambda_{1} = 1/4$, the negativity is $1/2$ whereas the bound is $(1/2)(\sqrt{2}-1) = 0.207$. 

Now let's look at Eq. (\ref{method2}) for the same state and for the same choice of $A$ and $B$.  In this case, $\| A^{\dagger}A\| = \|B^{\dagger}B\| = 1$.  The inequality becomes
\begin{equation}
\sqrt{\lambda_{0}\lambda_{1}} \leq \mu_{-} (1+\mu_{-})^{1/2} + \mu_{-}  ,
\end{equation}
and $N(\rho ) = \mu_{-}$.  We now have to try to solve this for $\mu_{-}$.  It is easier to solve a weaker inequality, using $(1+\mu_{-})^{1/2} \leq 1 + \mu_{-}/2$, 
\begin{equation}
\sqrt{\lambda_{0}\lambda_{1}} \leq \mu_{-}\left( \frac{\mu_{-}}{2} + 2 \right) ,
\end{equation}
or $(\mu_{-}^{2}/2) + 2\mu_{-} - \sqrt{\lambda_{0}\lambda_{1}} \geq 0$.  For this to be true $\mu_{-}$ must be greater than the two roots of the corresponding quadratic equation, which gives us
\begin{equation}
N(\rho) = \mu_{-} \geq \left( 4 + 2\sqrt{\lambda_{0}\lambda_{1}} \right) ^{1/2} -2 .
\end{equation}
Again when $\lambda_{0}\lambda_{1} = 0$, both the negativity and the bound are $0$, and when $\lambda_{0}\lambda_{1} = 1/4$, the bound is $\sqrt{5}-2 = 0.24$.

The previous example can be generalized to a system with noise. With noise, the 2 qubits system is in the mixed state
\begin{equation}
\rho = p  |\Psi\rangle \langle\Psi| + \frac{(1-p)}{4} I,\quad |\Psi\rangle_{ab} = \sqrt{\lambda_0}|01\rangle_{ab} + \sqrt{\lambda_1}|10\rangle_{ab}
\end{equation}
where $0\leq p \leq 1$ is the noise parameter. The negativity is found by diagonalizing $\rho^{T_B}$
\begin{equation}
N(\rho) = p\sqrt{\lambda_0\lambda_1+\frac{(1+p)}{4}} ,
\end{equation}
while $\kappa$ with noise is
\begin{equation}
\kappa=p^2 \lambda_0\lambda_1-\frac{(1-p)}{4} .
\end{equation}
These are plotted versus $\lambda_{0}$ in Figure 1 for the cases $p=1$ and $p=2/3$.  Since our bound is only useful when $\kappa > 0$, we see that in the presence of noise the lower bound is only useful for a limited range of $\lambda_{0}$.  In addition, its value is significantly reduced.
\begin{figure}[h]

\hspace*{-15mm}\begin{subfigure}{.55\linewidth}
  \includegraphics[width=\linewidth]{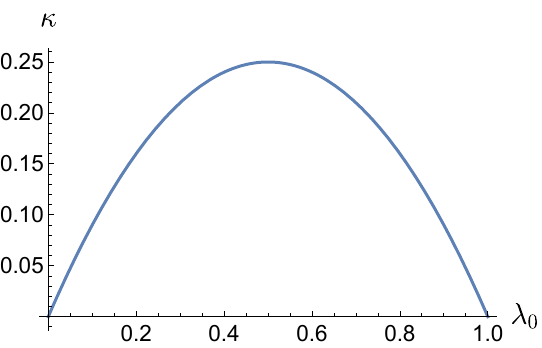}
  \caption{}
  \label{MLEDdet}
\end{subfigure}\hfill 
\begin{subfigure}{.55\linewidth}
  \includegraphics[width=1.4\linewidth]{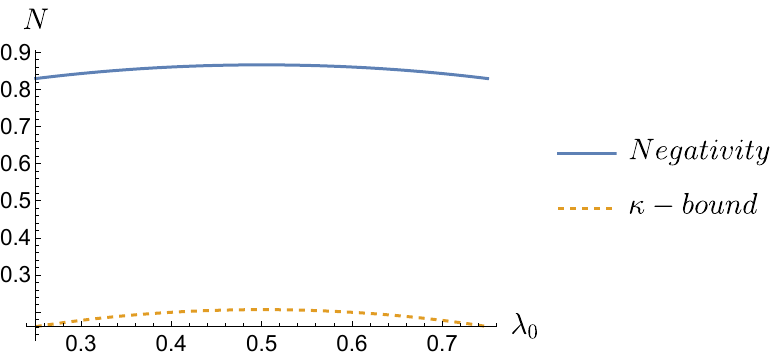}
  \caption{}
  \label{energydetPSK}
\end{subfigure}

\medskip 
\hspace*{-15mm}\begin{subfigure}{.55\linewidth}
  \includegraphics[width=\linewidth]{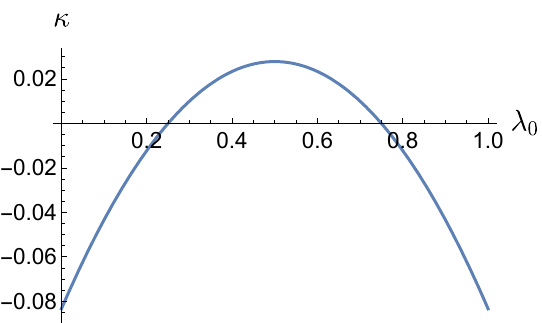}
  \caption{}
  \label{velcomp}
\end{subfigure}\hfill 
\begin{subfigure}{.55\linewidth}
  \includegraphics[width=1.4\linewidth]{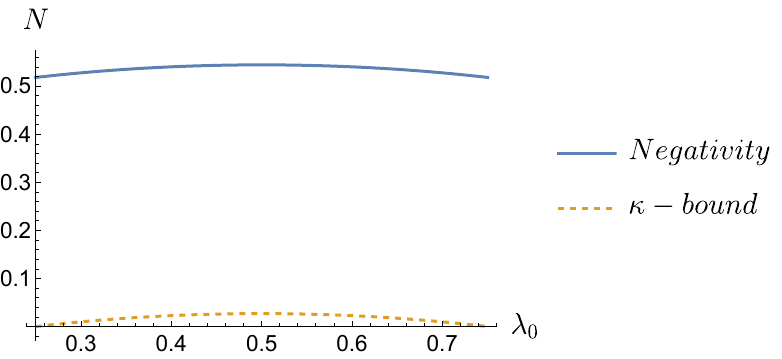}
  \caption{}
  \label{estcomp}
\end{subfigure}

\caption{(a),(b): $\kappa$ and Negativity/bound plot for $p=1$ (no noise).\\ \hspace*{1.4cm} (c),(d): $\kappa$ and Negativity/bound plot for $p=2/3$.}
\label{fig:roc}
\end{figure}

We can also go back to the first example and use a different set of operators $A,B$ in the hope that they will produce a better bound than  our original choice of $A_1=\sigma^{(-)}_{a}$, $B_1=\sigma^{(-)}_{b}$, which yielded $\kappa_1=\lambda_0\lambda_1$. The choice $A_2 =|+x\rangle \langle -x|$ and $B_2 =|-x\rangle \langle +x|$, where $|\pm x\rangle = (1/\sqrt{2}) (|0\rangle \pm |1\rangle )$, makes use of vectors that are superpositions of the Schmidt basis vectors $|0\rangle$ and $|1\rangle$, and this gives us

\begin{equation}
\kappa_2=\frac{1}{16}(1+2\sqrt{\lambda_0\lambda_1})^2 - \frac{1}{4}(1-2\sqrt{\lambda_0\lambda_1})
\end{equation}

and the new bound is

\begin{equation}
N(\rho) \geq \frac{1}{2} [ (1+4\kappa_2)^{1/2}-1]  
\end{equation}

\begin{figure}[h]
\hspace*{-30mm}\begin{minipage}{.75\textwidth}
  \includegraphics[width=.9\linewidth]{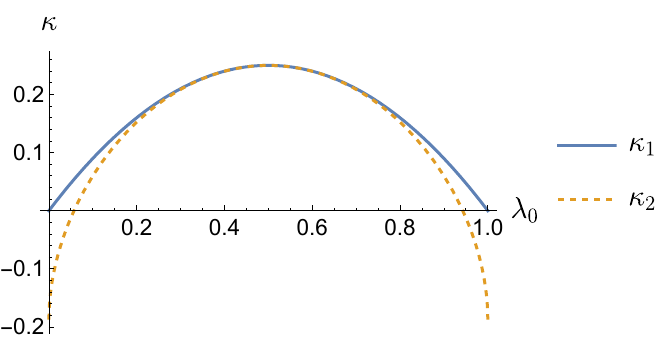}
  \captionof{figure}{$\kappa_1$, $\kappa_2$ as a function of $\lambda_0$. $\kappa_2$ becomes negative for certain values of $\lambda_0$, which numerically are found to be $0\leq\lambda_0\leq0.057109$ and $0.942891\leq\lambda_0\leq1$. These values are excluded in the negativity/bounds plot on the right.}
  \label{fig:test1}
\end{minipage}%
\hspace*{5mm}\begin{minipage}{.7\textwidth}
  \includegraphics[width=1.1\linewidth]{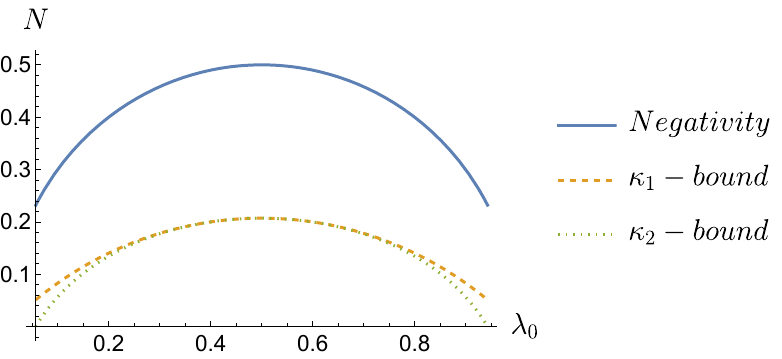}
  \captionof{figure}{Negativity and $\kappa$ bounds as a function of $\lambda_0$.  They are only plotted for the range in which the $\kappa$ bounds are greater than zero (see Fig.\ 2).  The $\kappa_1$-bound is better than the $\kappa_2$-bound.}
  \label{fig:test2}
\end{minipage}
\end{figure}

Bounds from both sets are plotted in Figure \ref{fig:test2} .  We see that the choice of operators derived from the Schmidt basis is slightly better than that derived from a superposition of these basis vectors.

The negativity is an entanglement monotone under LOCC, and this implies that if we trace out part of the system and find the negativity of the reduced system, it will be less than the negativity of the original system \cite{werner}.  Therefore, if we find a lower bound on the negativity of the reduced system, it is also a lower bound on the negativity of the original system.  With this in mind, let's now look at a four qubit system, where the first two qubits will be one subsystem and the second two will be the second subsystem, in the state
\begin{equation}
|\Psi\rangle = \sum_{j,k=0}^{1} \sqrt{\lambda_{jk}} |jj\rangle |kk\rangle .
\end{equation}
We will consider two choices of operators.  For the first, we will choose $A=|0\rangle\langle 1|\otimes I$ and $B=|1\rangle\langle 0| \otimes I$.  Doing so we find that
\begin{equation}
 \kappa = | \langle A^{\dagger}B\rangle |^{2} - \langle A^{\dagger}AB^{\dagger}B\rangle = (\sqrt{\lambda_{00}\lambda_{10}} + \sqrt{\lambda_{01}\lambda_{11}} )^{2} , 
\end{equation}
For the second choice we will choose two sets of operators and combine the results.  We first choose the  operators $A_1,B_1$ and $A_2,B_2$, where $A_1=|00\rangle\langle 10|$, $B_1=|10\rangle\langle 00|$, which yields $\kappa_{1}$ and then choose $A_2=|01\rangle\langle 11|$, $B_1=|01\rangle\langle 11|$, which yields $\kappa_{2}$. We have that
\begin{eqnarray}
&\kappa_1 = \lambda_{00} \lambda_{10} \nonumber \\
&\kappa_2 = \lambda_{01} \lambda_{11} .
\end{eqnarray}
We can then combine $\kappa_{1}$ and $\kappa_{2}$ into a new bound using Eq.\ (\ref{extend}).

The negativity of the state $|\Psi\rangle$ is
\begin{equation}
N(\rho) = \frac{1}{2}[(\sqrt{\lambda_{00}}+\sqrt{\lambda_{01}}+\sqrt{\lambda_{10}}+\sqrt{\lambda_{11}})^2 -1]
\end{equation}
If we fix $\lambda_{10}=\lambda_{00}$ and $\lambda_{01}=\lambda_{11}$, and we use the normalization condition of $|\Psi\rangle$, which is $2\lambda_{00} + 2\lambda_{11}=1$, we are able to compare the negativity with the $\kappa_1$-bound and the $\kappa_2$-bound by plotting everything as a function of $\lambda_{00}$ only. In this case, we have
\begin{eqnarray}
&&\kappa= \frac{1}{4},\,\kappa_1= \lambda_{00}^2,\,\kappa_2= (\frac{1}{2}-\lambda_{00})^2\\
&& N(\rho) = \frac{1}{2} (1+8\sqrt{\lambda_{00}\lambda_{11}})
\end{eqnarray}
and the plot is shown in Figure 4.  We note that  the bound obtained by combining $\kappa_1$ and $\kappa_2$ is lower than the $\kappa$-bound, showing that in this case the first choice of the operators $A$ and $B$ is the better one.

\begin{figure}
\centering
\includegraphics[width=\textwidth]{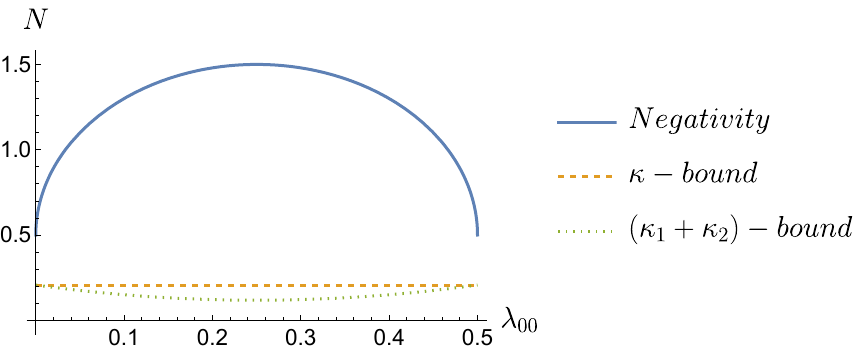}
\caption{Negativity and $\kappa$-bounds as a function of $\lambda_{00}$. The $\kappa$-bound is still better than $\kappa_1+\kappa_2$-bound.}
\end{figure}

\section{Pure states and the Schmidt basis}
Suppose we have a pure bipartite state $|\Psi\rangle_{ab}$, and we know some of the Schmidt basis vectors (an example of this is discussed in the Appendix).  In particular 
\begin{equation}
|\Psi\rangle_{ab} = \sum_{j=1}^{N}  \sqrt{\lambda_{j}} |u_{j}\rangle_{a} |v_{j}\rangle_{b} + |\Psi^{\prime}_{ab}\rangle .
\end{equation}
where $\{ |u_{j}\rangle_{a} \}$ and $\{ |v_{j}\rangle_{b} \}$ are subsets of the Schmidt basis, and $(\,_{a}\langle u_{j}|\,_{b}\langle v_{j}|)|\Psi^{\prime}\rangle = 0$ for $j=1,2, \ldots ,N$.  We would like to choose $A$ and $B$ so that $\langle A^{\dagger}A B^{\dagger}B\rangle = 0$.  One way to do this is to choose $K$ between $1$ and $N$ and define
\begin{eqnarray}
|\alpha\rangle = \sum_{j=1}^{K} |u_{j}\rangle & \hspace{5mm} & |\tilde{\alpha}\rangle = \sum_{j=K+1}^{N} |u_{j}\rangle \nonumber \\
|\beta\rangle = \sum_{j=1}^{K} |v_{j}\rangle & \hspace{5mm} & |\tilde{\beta}\rangle = \sum_{j=K+1}^{N} |v_{j}\rangle .
\end{eqnarray}
Then set $A = |\alpha\rangle\langle\tilde{\alpha}|$ and $B=|\tilde{\beta}\rangle\langle\beta |$.  This gives $A^{\dagger}A = |\tilde{\alpha}\rangle\langle\tilde{\alpha}|$ and $B^{\dagger}B = |\beta\rangle \langle\beta |$.  With this choice we have
\begin{eqnarray}
\langle A^{\dagger}AB^{\dagger}B\rangle & = & 0. \nonumber \\
\langle A^{\dagger}B\rangle & = & \left(\sum_{j=K+1}^{N} \sqrt{\lambda_{j}}\right) \left( \sum_{k =1}^{K} \sqrt{\lambda_{k}} \right) .
\end{eqnarray}
Note that if the above two equations hold, then 
\begin{equation}
\sum_{j=1}^{N} \sqrt{\lambda_{j}} \geq 2\sqrt{|\langle A^{\dagger}B\rangle |} .
\end{equation}
This gives for the negativity that
\begin{equation}
N(|\Psi\rangle_{ab}\langle\Psi |) \geq \frac{1}{2} \left( 4|\langle A^{\dagger}B\rangle |-1\right) .
\end{equation}
For this to be useful, we would need $|\langle A^{\dagger}B\rangle | > 1/4$.  Also note that we can choose $K$ to maximize $|\langle A^{\dagger}B\rangle |$.

As a simple example we can consider the case that $|\Psi_{ab}^{\prime}\rangle = 0$ and $\lambda_{j}=1/N$ for $j=1,2,\ldots ,N$, that is a maximally entangled state.  Assuming that $N$ is even and $K=N/2$, we find that the above inequality yields $N(|\Psi\rangle_{ab}\langle\Psi |) \geq \frac{1}{2} (N-1)$, which is, in fact, the negativity of the state.

\section{Conclusion}
We have shown that two conditions that can be used to detect entanglement in a state can also be used to provide a lower bound on the negativity of the state.  The conditions themselves are rather simple and can provide quick information about the negativity, whereas calculating the negativity itself involves diagonalizing the partial transpose of the density matrix.  In order to use the entanglement conditions, one needs to make a choice of operators, and this choice determines the negativity bounds one will obtain.  The effects of this choice were studied through a number of examples.

\section*{Acknowledgments}
This research was supported by NSF grant FET-2106447. The Flatiron Institute is supported by the Simons Foundation.

\section*{Appendix}
We want to provide an example of a situation in which one can know some of the Schmidt basis vectors of a model without having to solve the entire problem.  Consider a spin of size $j$ coupled to a single mode field by the Hamiltonian ($\hbar = 1$)
\begin{equation}
H=\omega S_{3}+ \omega a^{\dagger}a + g(S^{(+)} a + S^{(-)} a^{\dagger})  .
\end{equation}
The spin Hilbert space has the basis $|m\rangle$, where $-j \leq m \leq j$, and $S_{3}|m\rangle = m |m\rangle$.  The operators $S^{(\pm)}$ are the standard spin raising and lowering operators.  This Hamiltonian is used in quantum optics to describe the Dicke model in which a collection of $N$ two-level atoms ($j=N/2$) interacts with a single-mode field.  The number of excited atoms in the state $|m\rangle$ is $m+j$.  This system has a conservation law, the operator $S_{3}+a^{\dagger}a$ commutes with the Hamiltonian.

If we start in a state $|-j+l_{0}\rangle |0\rangle$, where $|0\rangle$ is the vacuum state of the field mode, the state at any later time can be expressed as
\begin{equation}
|\Psi\rangle = \sum_{l^{\prime}=0}^{l_{0}} d_{l^{\prime}}|-j+l^{\prime}\rangle |l_{0}-l^{\prime}\rangle .
\end{equation}
Note that this state is automatically expressed in its Schmidt basis as a result of the conservation law, so in this case we know all of the Schmidt vectors.  If we start in the superposition $(c_{0}|-j+l_{1}\rangle + c_{1}|-j+l_{2}\rangle ) |0\rangle$, where $l_{2}>2(l_{1}+1)$, then things are more complicated, but none the less we can identify some of the states in the Schmidt basis.  At a later time this state has the form
\begin{eqnarray}
|\Psi\rangle & = & c_{0} \left(  \sum_{l_{1}^{\prime}=0}^{l_{1}} d_{l_{1}^{\prime}}|-j+l_{1}^{\prime}\rangle |l_{1}-l^{\prime}_{1}\rangle \right) \nonumber \\
&& + c_{1}\left(  \sum_{l_{2}^{\prime}=0}^{l_{2}} f_{l^{\prime}_{2}}|-j+l^{\prime}_{2}\rangle |l_{2}-l^{\prime}_{2}\rangle \right) .
\end{eqnarray}
In order to find the Schmidt vectors, we find the reduced density matrices from $|\Psi\rangle\langle\Psi |$, one for the spin and one for the field, and find their eigenvectors.  From this, we find that the vectors $|-j+l_{1}+s\rangle |l_{2}-l_{1}-s\rangle$, where $1\leq s \leq l_{2} - 2l_{1} -1$ are Schmidt vectors.


\begin{thebibliography}{99}
\bibitem{z1} K.~Życzkowski, P.~Horodecki,  A.~Sanpera, and M.~Lewenstein, Phys.\ Rev.\ A {\bf 58}, 883 (1998).
\bibitem{z2} K.~Życzkowski, Phys.\ Rev. A {\bf 60},  3496 (1999).
\bibitem{werner} G.~Vidal and R.~F.~Werner, Phys.\ Rev.\ A {\bf 65}, 032314 (2002).
\bibitem{cavalcanti} D.~Cavalcanti and M.~O. Terra Cunha, Appl.\ Phys.\ Lett. {\bf 89}, 084102 (2006).
\bibitem{plenio} K.~N.~R.~ Audenaert and M.~B.~Plenio, New J.\ Phys. {\bf 8}, 266 (2006).
\bibitem{eisert} J. Eisert, F.G.S.L. Brandao, and K.M.R. Audenaert, New J. Phys. {\bf 9}, 46 (2007).
\bibitem{guhne} O.~G\"{u}hne, M.~Reimpell, and R.~F.~Werner, Phys.\ Rev.\ Lett. {\bf 98}, 110502 (2007).
\bibitem{hillery1} M.~Hillery and M.~S.~Zubairy, Phys.\ Rev.\ Lett. {\bf 96}, 050503 (2006).
\bibitem{hillery2} M.~Hillery and M.~S.~Zubairy, Phys.\ Rev.\ A {\bf 74}, 032333 (2006).
\bibitem{hillery3}M.~Hillery, H.~T.~Dung, and J.~Niset, Phys.\ Rev.\ A {\bf 80}, 052335 (2009).
\bibitem{hillery4} M.~Hillery, H.~T.~Dung, and H.~Zheng, Phys.\ Rev.\ A {\bf 81}, 062322 (2010).
\bibitem{wolk} S.~W\"{o}lk, M.~Huber, and O.~G\"{u}hne, Phys.\ Rev.\ A {\bf 90}, 022315 (2014).
\end{thebibliography}
\end{document}